\documentclass[twocolumn,prl,preprintnumbers]{revtex4}
\usepackage{graphicx}
\usepackage{dcolumn}
\usepackage{bm}

\newcommand{\eps}{\epsilon}
\newcommand{\epsp}{\epsilon^\prime}

\newcommand{\bea}{\begin{eqnarray}}
\newcommand{\eea}{\end{eqnarray}}
\newcommand{\beq}{\begin{equation}}
\newcommand{\eeq}{\end{equation}}
\newcommand{\nn}{\nonumber}
\newcommand{\nl}{\nonumber\\}

\newcommand{\real}{{\rm Re}}
\newcommand{\imag}{{\rm Im}}

\newcommand{\PL}[3]{{Phys. Lett.} {\bf#1,} {#3} {(#2)}} 
 
\newcommand{\PR}[3]{{Phys. Rev.} {\bf#1,} {#3} {(#2)}} 
\newcommand{\NP}[3]{{Nucl. Phys.} {\bf#1,} {#3} {(#2)}} 
\newcommand{\EPJ}[3]{{Eur. Phys. J.} {\bf#1,} {#3} {(#2)}} 
 
\newcommand{\cO}{{\cal O}}
\newcommand{\cL}{{\cal L}}

\newcommand{\cA}{{\cal A}}

\newcommand{\ba}{\begin{array}{c}}
\newcommand{\bat}{\begin{array}{cc}}
\newcommand{\ea}{\end{array}}

\def\slashchar#1{\setbox0=\hbox{$#1$}\dimen0=\wd0%
\setbox1=\hbox{/}\dimen1=\wd1%
\ifdim\dimen0>\dimen1%
\rlap{\hbox to
\dimen0{\hfil/\hfil}}#1\else                                     
\rlap{\hbox to \dimen1{\hfil$#1$\hfil}}/\fi}
\newcommand{\RE}{\mbox{\rm Re}} 
 
\begin{document}

\preprint{IFIC/03-29}
\preprint{UWThPh-2003-2}

\title{
Isospin violation in $\epsp$
}
\thanks{
Work supported in part by HPRN-CT2002-00311 (EURI\-DICE) and by
Acciones Integradas, Project No. 19/2003 (Austria),
HU2002-0044 (MCYT, Spain).
}
\author{V. Cirigliano}
\author{A. Pich}%
\affiliation{
Departament de F\'{\i}sica Te\`orica, IFIC, CSIC --- 
Universitat de Val\`encia \\ 
Edifici d'Instituts de Paterna, Apt. Correus 22085, E-46071 
Val\`encia, Spain 
}
\author{G. Ecker}
\author{H. Neufeld}
\affiliation{
Institut f\"ur Theoretische Physik, Universit\"at 
Wien\\ Boltzmanngasse 5, A-1090 Vienna, Austria 
}%

\date{\today}

\begin{abstract}
On the basis of a next-to-leading-order calculation in chiral
perturbation theory, the first complete analysis of isospin
breaking for direct CP violation in $K^0 \to 2 \pi$ decays is
performed. We find a destructive interference between three different
sources of isospin violation in the CP violation parameter $\epsp$.
Within the uncertainties of large-$N_c$ estimates for the low-energy 
constants, the isospin violating correction for $\epsp$ is below 15 $\%$.
\end{abstract}

\pacs{Valid PACS appear here}
\maketitle

Direct CP violation in $K^0 \to 2\pi$ decays is now 
established \cite{nir02} to an accuracy of less than 10 $\%$:
\begin{equation} 
\RE ~\epsp/\eps = (1.66 \pm 0.16)\times 10^{-3}~.
\end{equation}
The theoretical situation is less satisfactory due to large
uncertainties in the hadronic matrix elements of
four-quark operators. Although several calculations exist that come
close to the experimental result (e.g., Ref.~\cite{pps01}) the
accuracy is still rather limited.

$K$ decays involve a delicate interplay between electro\-weak and strong
interactions in the confinement regime. Chiral perturbation theory
(CHPT) provides a convenient framework for a systematic low-energy
expansion of the relevant amplitudes. In this letter,
we concentrate on a quantitative analysis of isospin
violation in the CP-violating parameter $\epsp$. The results presented
here are part of a complete calculation of $K \to 2 \pi$
decays \cite{cenp03c} to next-to-leading order in the chiral expansion 
and to first order in isospin violation, including both strong isospin 
violation ($m_u \neq m_d$) and electromagnetic corrections.

Although isospin violation is in general a small effect
it must be included 
\cite{etaetap,emnp00,cdg00,wm01} in a precision calculation of
$\epsp$ because it affects the destructive interference between the
two main contributions to $\epsp$ from normal and electromagnetic 
penguin operators. 

The main features of our approach are the following:
\paragraph{1.}
We include for the first time both strong and
electromagnetic isospin violation in a joint analysis.
\paragraph{2.}
Nonleptonic weak amplitudes in CHPT depend on a number of low-energy
constants (LECs): we use leading large-$N_c$ estimates for those
constants \cite{pps01,cenp03c}. 
Uncertainties arise from (i) input parameters in the leading $1/N_c$
expressions as well as from (ii) potentially large subleading effects 
in $1/N_c$.  We discuss the impact of both (i) and (ii) on the 
relevant quantities.

We adopt the following parametrization of $K \to \pi\pi$ amplitudes 
\cite{cenp03c,cdg00}:
\begin{eqnarray}
A(K^0 \to \pi^+ \pi^-) &=& 
A_{0} \, e^{i \chi_0}  + { 1 \over \sqrt{2}} \,   A_{2}\,  e^{i\chi_2 }
 \nn \\
A(K^0 \to \pi^0 \pi^0) &=& 
A_{0} \, e^{i \chi_0}  - \sqrt{2} \,   A_{2}\,  e^{i\chi_2 }  
\label{eq:amps}
\\ 
A(K^+ \to \pi^+ \pi^0)   &=& {3 \over 2} \,  
A_{2}^{+} \,  e^{i\chi_2^{+}}  ~. 
\nonumber
\end{eqnarray}
In the limit of CP conservation, the amplitudes $A_{0}, A_{2}$, and  
$A_{2}^+$ are real and positive. In the isospin limit,
$A_{2}=A_{2}^+$, $\chi_2=\chi_2^{+}$ in the standard model and the 
phases $\chi_i$ coincide with the corresponding pion-pion phase shifts 
at $E_{\rm cms}=M_K$. In terms of these amplitudes, the direct CP 
violation parameter $\epsp$ is given by
\begin{equation} 
\epsp  = - \displaystyle\frac{i}{\sqrt{2}} \, e^{i ( \chi_2 - \chi_0 )} \, 
\displaystyle\frac{\real A_{2}}{ \real A_{0}} \,  
\left[ 
\displaystyle\frac{\imag A_{0}}{ \real A_{0}} \, - \,  
\displaystyle\frac{\imag A_{2}}{ \real A_{2}} \right]  
\ . 
\label{eq:cp1}
\eeq

For the systematic low-energy expansion of $K \to \pi\pi$ amplitudes
to next-to-leading order, a number of effective chiral Lagrangians is
needed. Here, we only write down the relevant nonleptonic
weak Lagrangian, referring to Ref.~\cite{cenp03c} for further details:
\begin{eqnarray} 
\cL_{\rm weak} &=&  G_8  F^4  ( L_{\mu} L^\mu)_{23} 
+  G_8 g_{\rm ewk} e^2 F^6 (U^\dagger Q U)_{23} 
\nl
& + & G_{27} F^4 \left( L_{\mu 23} L^\mu_{11} + 
{2\over 3} L_{\mu 21} L^\mu_{13}\right) \label{eq:Lweak} \\
& + &   \sum_i\;G_8 N_i  F^2 O^8_i +
  \sum_i\; G_{27} D_i  F^2 O^{27}_i \ + \ {\rm h.c.}\nn
\end{eqnarray} 
The matrix $L_{\mu}=i U^\dagger D_\mu U$  represents the octet of
$V-A$ currents to lowest order in derivatives where the $SU(3)$ matrix
field $U$ contains the pseudoscalar fields. 
$Q$ represents the light quark charge matrix. 
In addition to the 
lowest-order couplings $G_8, G_{27}$, and $G_{8} g_{\rm ewk}$, 
this Lagrangian contains 
LECs $G_8 N_i$, $G_{27} D_i$ of $\cO(G_F p^4)$ \cite{GFp4}. Due to
the disparity in size between $G_8$ and $G_{27}$ ($\Delta I=1/2$ rule),
we consider isospin violating effects in the octet part only. More
precisely, we work up to $\cO(G_8 p^4,G_8 (m_u-m_d) p^2,G_8 e^2 p^2)$ 
and to $\cO(G_{27} p^4)$ for octet and 27-plet amplitudes, respectively. 
For this purpose, the chiral effective Lagrangian for strong interactions to
$\cO(p^4)$, the electromagnetic Lagrangian to $\cO(e^2 p^2)$ and the
electroweak Lagrangian to $\cO(G_8 e^2 p^2)$ are also needed
\cite{cenp03c}.   
 
The resulting decay amplitudes depend on a large number of
renormalized LECs. In addition to the well-known strong LECs of
$\cO(p^4)$, we make use of existing estimates for the electromagnetic
couplings of $\cO(e^2 p^2)$ \cite{e2p2}. For the nonleptonic weak
LECs of $\cO(G_F p^4)$ and $\cO(G_8 e^2 p^2)$, we employ leading
large-$N_c$ estimates \cite{pps01,cenp03c}. 
The starting point is the effective
$\Delta S=1$ Hamiltonian in the three-flavour theory,
\begin{equation} 
\label{eq:Heff}
 {\cal H}_{\mathrm eff}^{\Delta S=1}= \displaystyle\frac{G_F}{\sqrt{2}}
 V_{ud}^{\phantom{*}}\,V^*_{us}\,  \sum_i  C_i(\mu_{\rm SD}) \; 
Q_i (\mu_{\rm SD}) \; ,
\end{equation} 
obtained after integrating out all fields with masses larger than 
$\mu_{\rm SD} \simeq 1$ GeV. The local four-quark operators $Q_i$ contain
only the light degrees of freedom whereas the Wilson coefficients 
$C_i$ are functions of heavy masses and CKM parameters~\cite{munich,rome}:
\begin{eqnarray} 
C_i (\mu_{\rm SD}) &=&  z_i (\mu_{\rm SD}) \ + \  \tau \, 
y_i (\mu_{\rm SD}) \\
\tau &=&  - \displaystyle\frac{V_{td}^{\phantom{*}}\,
V^*_{ts}}{V_{ud}^{\phantom{*}}\,V^*_{us}} \ . 
\end{eqnarray} 
All CP-violating quantities are proportional to $\imag ~\tau$. In our
analysis, only ratios of CP-violating amplitudes will appear so we do
not need an explicit value for $\tau$. 

At leading order in $1/N_c$, the matching
between the three-flavour quark theory and CHPT can be done exactly
because the T-product of two colour-singlet quark currents factorizes.
Since quark currents have well-known realizations in CHPT,
the hadronization of the weak operators $Q_i$ is 
straightforward \cite{pps01,cenp03c}.
The weak LECs are expressed in terms of strong LECs of order 
$p^4$, $p^6$, and $e^2 p^2$.
The $\cO (p^4)$ and $\cO(p^6)$ couplings arising from bosonization of
$Q_6$ are estimated within resonance saturation and are determined by
scalar exchange (for details see Refs.~\cite{cenp03c,cenp03a}).
Although admittedly this is not a complete estimate for the $\cO(p^6)$
couplings it certainly provides the correct order of magnitude.

Subleading effects in $1/N_c$ are known to be sizable  
at leading chiral order. 
We will therefore not use the large-$N_c$ values for $\real G_8$,
$\real G_{27}$ in the numerical analysis but instead determine these
couplings from a fit to the $K \to \pi\pi$ branching ratios
\cite{cenp03c}.  The other combination of interest is the ratio $\imag
(G_8 g_{\rm ewk})/\imag G_8$. In this case the size of $1/N_c$
effects is approximately determined 
by existing calculations going beyond factorization~\cite{nonfact}.
It turns out that corrections to this ratio are not very large
(roughly $-30 \% $).  The dominant uncertainty comes then from the
input parameters in the factorized expressions.

At next-to-leading chiral order we estimate subleading effects in
$1/N_c$ by varying the chiral renormalization scale at which the
large-$N_c$ results are supposed to apply.  We consider two options
for the higher-order couplings $G_8 N_i$, \dots ~in the Lagrangian
(\ref{eq:Lweak}): one may either adopt the large-$N_c$ predictions for
the ratios $(G_8 N_i)/G_8$ or directly for the couplings $G_8 N_i$.  The
uncertainties related to these two choices will be taken into account.
We also include in the errors the effect of changing the
short-distance renormalization scale $\mu_{\rm SD}$ between $M_\rho$
and 1.3 GeV, with the central value at 1 GeV. 
 
Whenever the errors are sizable, they are however largely 
dominated by the intrinsic uncertainty (an effect of higher order in
$1/N_c$) at which chiral scale $\nu_{\rm CH}$ the large-$N_c$
estimates for the renormalized LECs $G_8 N_i^r(\nu_{\rm CH})$, \dots
~apply. We vary this scale between 0.6 GeV and 1 GeV, with the
central values at $M_\rho$.

We now turn to the different sources of isospin violation in $\epsp$. 
The expression (\ref{eq:cp1}) is valid to first order in CP violation.
Since $\imag A_{I}$ is CP odd the quantities $\real A_{I}$ and 
$\chi_{I}$ are only needed in the CP limit ($I=0,2$). We disregard the 
phase that can be obtained from the $K \to \pi\pi$ branching ratios. 
The same branching ratios are usually employed to extract the ratio 
$\omega_S=\real A_{2}/ \real A_{0}$ assuming isospin conservation. 
Accounting for isospin violation via the general parametrization
(\ref{eq:amps}), one is then really evaluating 
$\omega_+ = \real A_{2}^{+}/\real A_{0}$ rather than $\omega_S$.
The two differ by a pure $\Delta I=5/2$ effect:
\bea
\omega_S &=& \omega_+ \, \left( 1 +  f_{5/2} \right) \\
f_{5/2} &=& \displaystyle\frac{\real A_{2}}{\real A^{+}_{2}} - 1  
 \ .   
\label{cp4}
\eea
Because $\omega_+$ is directly related to branching ratios it proves
useful to keep $\omega_+$ in the normalization of $\epsp$, introducing 
the $\Delta I=5/2$ correction $f_{5/2}$ \cite{cdg00}.

Since $\imag A_2$ is already first order in isospin violation
the formula for $\epsp$ takes the following form, with all first-order
isospin breaking corrections made explicit:
\begin{eqnarray}  
\epsp =  - \displaystyle\frac{i}{\sqrt{2}}  e^{i ( \chi_2 - \chi_0 )}  
\omega_+  & \left[ \displaystyle\frac{\imag A_{0}^{(0)}}{ \real A_{0}^{(0)}} 
(1 + \Delta_0 + f_{5/2}) \right. \nl
 & \left. \hspace*{1cm} - \displaystyle\frac{\imag A_{2}}
{\real A_{2}^{(0)}} \right] ~, 
\label{eq:cpiso}
\end{eqnarray} 
where 
\begin{equation} 
\Delta_0 = \displaystyle\frac{\imag A_0}{\imag A_0^{(0)}} \cdot 
\displaystyle\frac{\real A_0^{(0)}}{\real A_0}  - 1  
\end{equation}
and the superscript $(0)$ denotes the isospin limit.

To isolate the isospin breaking corrections in $\epsp$, we write the
amplitudes $A_0,A_2$ more explicitly as
\begin{eqnarray}  
A_0 \,e^{i \chi_0 } = \cA_{1/2} &=&  \cA_{1/2}^{(0)} + \delta \cA_{1/2} 
\nn \\
 A_2 \,e^{i \chi_2 }= \cA_{3/2} &=&  \cA_{3/2}^{(0)} + \delta \cA_{3/2}+ 
\cA_{5/2} ~,
\label{eq:A02}   
\end{eqnarray}
where $\delta \cA_{1/2,3/2}$, $\cA_{5/2}$ are first order in isospin
violation. To the order we are considering, the amplitudes
$\cA_{\Delta I}$ have both
absorptive and dispersive parts. To disentangle the (CP conserving)
phases generated by the loop amplitudes from the CP-violating phases
of the various LECs, we express our results explicitly in terms of 
${\cal D}isp ~\cA_{\Delta I}$ and ${\cal A}bs ~\cA_{\Delta I}$. 

To first order both in CP violation and in isospin breaking, we obtain
\begin{widetext}
\begin{eqnarray}  
\Delta_0 &=& -2 \left| \cA_{1/2}^{(0)} \right|^{-2} \, 
\left( \real  [ {\cal D}isp \, \cA_{1/2}^{(0)} ]~ 
\real [{\cal D}isp \, \delta \cA_{1/2}] + 
\real [ {\cal A}bs \,\cA_{1/2}^{(0)}]~ 
\real [{\cal A}bs \, \delta \cA_{1/2}]
\right) \nl
&+& \left[ \imag [ {\cal D}isp \,\cA_{1/2}^{(0)} ] \,
\real [ {\cal D}isp \, \cA_{1/2}^{(0)} ] + 
\imag [ {\cal A}bs \,\cA_{1/2}^{(0)}]  \,
\real[ {\cal A}bs \,  \cA_{1/2}^{(0)} ] \right]^{-1} 
\left\{\imag [ {\cal D}isp \,\delta\cA_{1/2} ] \,
\real [ {\cal D}isp \,\cA_{1/2}^{(0)} ] \right. \nn  \\ 
&& + \left.
\imag [ {\cal D}isp \,\cA_{1/2}^{(0)} ] \,
\real[ {\cal D}isp \,  \delta\cA_{1/2} ] 
+   \imag [  {\cal A}bs \,\delta\cA_{1/2} ] \,
\real[ {\cal A}bs \,  \cA_{1/2}^{(0)} ] + 
\imag [ {\cal A}bs \,\cA_{1/2}^{(0)} ] \,
\real [ {\cal A}bs \, \delta\cA_{1/2} ] \right\} \label{eq:D0} \\ 
f_{5/2} &=& \displaystyle\frac{5}{3} \left| \cA_{3/2}^{(0)} \right|^{-2} \, 
\left\{\real [ {\cal D}isp \,\cA_{3/2}^{(0)} ] \,
\real [ {\cal D}isp \, \cA_{5/2} ] + 
\real [ {\cal A}bs \,\cA_{3/2}^{(0)}]  \,
\real[ {\cal A}bs \,  \cA_{5/2}^{(0)} ] \right\}
\label{eq:f52} \\
\imag A_2 \ \,  &=&  \left| \cA_{3/2}^{(0)} \right|^{-1} \, 
\left\{\imag [ {\cal D}isp \, 
\left( \delta \cA_{3/2} + \cA_{5/2} \right) ]  \,
\real [ {\cal D}isp \, \cA_{3/2}^{(0)} ] 
 + \imag [ {\cal A}bs \left( \delta  \cA_{3/2} + \cA_{5/2} \right) ]  \,
\real[ {\cal A}bs \,  \cA_{3/2}^{(0)} ] \right\}\label{eq:ima2} ,   
\end{eqnarray}
\end{widetext}
where $| \cA_n^{(0)} | =  
\sqrt{ (\real [{\cal D}isp \,  \cA_n^{(0)}  ])^2 +
(\real [{\cal A}bs \,  \cA_n^{(0)} ])^2 } $.
 
These expressions are general results to first order in CP and isospin
violation but they are independent of the chiral expansion. Working
strictly to a specific chiral order, these formulas 
simplify. We prefer to keep them in their general form
but we will discuss later the numerical differences between the
complete and the systematic chiral expressions. The differences are
one indication for the importance of higher-order chiral corrections.

Although $\imag A_2$ is itself first order in isospin breaking we now
make the usual (but scheme dependent) separation of the
electroweak penguin contribution to $\imag A_2$ from the
isospin breaking effects generated by other four-quark operators: 
\begin{equation} 
\imag A_2 = \imag A_2^{\rm emp} \ + \ \imag  A_2^{\rm non-emp} \ .  
\end{equation}    
In order to perform such a separation within the CHPT approach,
we need to identify the electroweak penguin contribution to a given
low-energy coupling.  In other words, we need a matching procedure
between CHPT and the underlying theory of electroweak and strong
interactions.  Such a matching procedure is given here by working at
leading order in $1/N_c$.  
Then, the electroweak LECs of $\cO(G_8 e^2 p^n)$ ($n=0,2$) in $\imag
A_2^{\rm non-emp}$ must be calculated by setting to zero the Wilson
coefficients $C_7,C_8,C_9,C_{10}$ of electroweak penguin operators.

Splitting off the electromagnetic penguin contribution to $\imag A_2$
in this way, we can now write $\epsp$ in a more familiar form as
\begin{equation}  
\epsp = - \displaystyle\frac{i}{\sqrt{2}} \, e^{i ( \chi_2 - \chi_0 )} \, 
\omega_+  \,   \left[ 
\displaystyle\frac{\imag A_{0}^{(0)} }{ \real A_{0}^{(0)} } \, 
(1 - \Omega_{\rm eff}) - \displaystyle\frac{\imag A_{2}^{\rm emp}}{ \real
  A_{2}^{(0)} } \right]  
\label{eq:cpeff}
\end{equation} 
where 
\begin{eqnarray}  
\Omega_{\rm eff} &=& \Omega_{\rm IB} - \Delta_0 - f_{5/2}  
\label{eq:omegaeff} \\
\Omega_{\rm IB} &=& \displaystyle\frac{\real A_0^{(0)} }
{ \real A_2^{(0)} } \cdot \displaystyle\frac{\imag A_2^{\rm non-emp} }
{ \imag A_0^{(0)} } ~.
\end{eqnarray}  
The quantity $\Omega_{\rm eff}$ includes all effects to leading order
in isospin breaking and it generalizes the more traditional parameter
$\Omega_{\rm IB}$. Although $\Omega_{\rm IB}$ is in principle
enhanced by the large ratio $\real A_0^{(0)}/ \real A_2^{(0)}$ the
actual numerical analysis shows all three terms in
(\ref{eq:omegaeff}) to be relevant when both
strong and electromagnetic isospin violation are included.

We present numerical results for the following two cases:
\begin{enumerate} 
\item[i.] We calculate $\Omega_{\rm eff}$ and its components for 
  $\alpha=0$ (purely strong isospin violation). In this case, there is 
  a clean separation of isospin violating effects in $\imag A_{2}$.
  We compare the lowest-order result of $\cO(m_u - m_d)$ with the full 
  result of  $\cO[(m_u - m_d)p^2]$.
\item[ii.] Here we include electromagnetic corrections explicitly, comparing
  again $\cO(m_u - m_d,e^2 p^0)$ with $\cO[(m_u - m_d)p^2,e^2 p^2]$. In
  this case, the splitting between $\imag A_2^{\rm emp}$ and 
  $\imag  A_2^{\rm non-emp}$ is performed at leading order in
  $1/N_c$.
\end{enumerate} 

In Table \ref{tab:tab1} the uncertainties in the ``LO'' entries 
are dominated by error propagation from the input parameters 
(these ranges account however for known $1/N_c$ corrections). 
Apart from loops, new effects in ``NLO'' entries depend on ratios of
next-to-leading to leading-order LECs. In Table \ref{tab:tab1}, we use
the leading $1/N_c$ estimates for the ratios $(G_8 N_i)/G_8$, \dots , 
and estimate the uncertainty as discussed above. 
The final error for each of the quantities $\Omega_{\rm IB}$,
$\Delta_0$, $f_{5/2}$, and $\Omega_{\rm eff}$ is obtained by adding in
quadrature the LO error and the one associated to weak LECs at NLO. 
Moreover, only $f_{5/2}$ and $\real A_0^{(0)} / \real A_2^{(0)}$
depend on the ratio $G_8/G_{27}$. In these cases we rely on the
phenomenological value implied by our fit \cite{cenp03c}.
Some of the errors in Table \ref{tab:tab1} are manifestly
correlated, e.g., in the LO column for $\alpha \neq 0$.

\begin{table}[ht]
\begin{center}
\begin{tabular}{|c|cc|cc|}
\hline
 & & & & \\
 & \multicolumn{2}{c|}{ $\alpha=0$}& \multicolumn{2}{c|}{ $\alpha \neq
 0$}  \\[5pt]
 &  \mbox{     } LO \mbox{     }  & \mbox{     } NLO \mbox{     } & 
\mbox{     }  LO \mbox{     } & \mbox{     } NLO \mbox{     } \\[5pt]
\hline
 & & & & \\
$\Omega_{\rm IB}$ & $11.7$ & $15.9 \pm 4.5$ & $ 18.0 \pm 6.5 $  & 
$ 22.7  \pm  7.6 $ \\[5pt]
$\Delta_0$ & $- 0.004$ & $- 0.41 \pm 0.05$ & $8.7 \pm 3.0$ & 
$ 8.3  \pm 3.6 $ \\[5pt]
$f_{5/2}$ & $0$ & $0$ & $ 0  $ & 
$ 8.3  \pm 2.4 $ \\[10pt]
\hline 
 & & & & \\
\mbox{   } $\Omega_{\rm eff}$ \mbox{   } & $11.7$ & $16.3 \pm 4.5$ & 
$9.3 \pm  5.8 $   &  $ 6.0  \pm  7.7$  \\[10pt]
\hline
\end{tabular}
\end{center}
\caption{Isospin violating corrections for $\epsp$ in units of
  $10^{-2}$. The first two columns refer to strong isospin violation
  only ($m_u \neq m_d$), the last two contain the complete results
  including electromagnetic corrections. LO and NLO denote leading and
  next-to-leading orders in CHPT.}
\label{tab:tab1}
\end{table}

The NLO results are obtained with the full expressions (\ref{eq:D0}), 
\dots, (\ref{eq:ima2}). Using instead the simplified expressions
corresponding to a fixed chiral order, the results are found
to be well within the quoted error bars. The same is true for the
alternative procedure of applying large-$N_c$ directly to the LECs 
$G_8 N_i$, \dots. The isospin violating ratios in Table \ref{tab:tab1} 
are hardly changed at all in this case. Altogether, we expect our errors 
to be realistic estimates of higher-order effects in the chiral expansion.

As first observed by Gardner and Valencia \cite{gv99}, there is a
source of potentially large isospin violation through terms of the type
$(m_u - m_d) G_8 N_i$, which get contributions from strong LECs of $\cO(p^6)$
at leading order in large $N_c$. The crucial LEC of $\cO(p^6)$ can be
related \cite{abt00} to the mass splitting in the lightest scalar
nonet that survives in the large-$N_c$ limit. The results in Table 
\ref{tab:tab1} are based on the evidence
discussed in Ref.~\cite{cenp03a} that the isotriplet and isodoublet
states in the lightest scalar nonet are $a_0(1450)$ and $K_0^{*}(1430)$,
respectively (Scenario A). In the large-$N_c$ limit, the results are 
independent of the assignment of isosinglet scalar resonances.
If instead the $a_0(980)$ is the isotriplet member of the lightest
nonet surviving in the large-$N_c$ limit
the couplings $(m_u - m_d) G_8 N_i$ would be strongly
enhanced \cite{gv99}, implying large (of the order of 100 $\%$) and 
negative values of $\Omega_{\rm IB}$. In this case, $\epsp/\eps$
would be more than twice as big as in the isospin limit. We consider 
this an additional argument against such a scenario for the lightest 
scalar nonet (Scenario B in Ref.~\cite{cenp03a}).

Finally, we have investigated the impact of some subleading effects
along the lines of Ref.~\cite{cenp03a}. Although by no means a
systematic expansion in $1/N_c$, those nonet breaking terms may
furnish yet another indication for the intrinsic uncertainties of some
of the LECs. The size of those terms depends on the assignment of
isosinglet scalar resonances.  Since nonet breaking effects are large
in the scalar sector they affect most of the entries in Table
\ref{tab:tab1} in a non-negligible way, although always within the
quoted uncertainties.  Employing again scenario A for the lightest scalar
nonet \cite{cenp03a}, $\Omega_{\rm eff}$ in (\ref{eq:cpeff}) decreases
from $6.0 \cdot 10^{-2}$ to $-1.4 \cdot 10^{-2}$.

We have performed a complete analysis of isospin
violation in $\epsp$, due to both the light quark mass
difference and electromagnetism. In particular, we have included
for the first time isospin violation in the ratio 
$\imag A_0 / \real A_0$, parametrized by the quantity $\Delta_0$. This
ratio gets only a small contribution from strong isospin violation but
the electromagnetic part is important, being dominated by
electromagnetic penguin contributions. Both
$\Delta_0$ and the purely electromagnetic $\Delta I=5/2$ contribution
$f_{5/2}$ interfere destructively with $\Omega_{\rm IB}$ to yield a
final value $\Omega_{\rm eff}= (6.0 \pm 7.7) \cdot 10^{-2}$ for the
overall measure of isospin violation in $\epsp$. 
If electromagnetic penguin contributions
are included in theoretical calculations of $\imag A_0 / \real A_0$,
$\Delta_0$ can be dropped in $\Omega_{\rm eff}$ to
a very good approximation. Finally, if all electromagnetic corrections 
are included in $\imag A_0 / \real A_0$, $\imag A_2 / \real
A_2$, and $\real A_2 / \real A_0$, $\Omega_{\rm eff}$
is essentially determined by $\Omega_{\rm IB}$.

\begin{acknowledgments}
We thank J. Bijnens for useful information.
The work of V.C. and A.P. has been supported in part by
MCYT, Spain (Grant No. FPA-2001-3031) and by ERDF funds from the
European Commission.

\end{acknowledgments}

\end{document}